\documentclass[12pt]{article}
\usepackage{amsmath}
\usepackage{slashed}
\usepackage{tikz}
\usepackage{tikz-feynman}
\usepackage{tikz-feynhand}
\usepackage{float}
\usepackage{graphicx}
\usepackage[left=0.8in,right=0.8in,%
top=1in,bottom=1in]{geometry}
\begin{document}

\begin{titlepage}

\begin{center}
{\bf\Large Strange quark mass effect in $B_s\to\gamma\gamma,\gamma \ell\bar{\ell}$ decays  }
\end{center}
\vspace{0.5cm}

\begin{center}
{\bf Dong-Hao Li$^{a,b}$, Lei-Yi Li$^{a,b}$, Cai-Dian L\"u$^{a,b}$ and Yue-Long Shen$^c$\footnote{Email: shenylmeteor@ouc.edu.cn, corresponding author}}\\ \vspace{0.5cm}
{\sl $^a$ Institute of High Energy Physics, CAS, P.O. Box 918(4) Beijing 100049,  China }\\
{\sl $^b$ \, School of Physics, University of Chinese Academy of Sciences, Beijing 100049, China }\\
{\sl $^c$ College of Physics and Photoelectric Engineering, Ocean University of China, Qingdao 266100,  China }\\
\end{center}
\vspace{0.2cm}

\begin{abstract}
In this paper we investigate  the next-to-leading power contribution  to the $B_{ \, s} \to \gamma \gamma$ and $B_{ \, s} \to \gamma \ell\bar{\ell}$ decays from the strange quark mass effect with the dispersion approach which is QCD inspired and more predictive. We have presented the analytic expression of the quark mass contribution  in the $B_{ \, s} \to \gamma \gamma$ and $B_{ \, s} \to \gamma \ell\bar{\ell}$ decays, together with a new term that is missed in the previous study. The numerical results of the strange quark mass contribution to the $B_{ \, s} \to \gamma \gamma$ decay is about $6\%$  relative to the total branching ratio, while it is relatively small in the $B_ { \, s} \to \gamma \ell\bar{\ell}$ decay due to the large resonance contribuiton.

\end{abstract}

\vfil

\end{titlepage}

\section{Introduction}

The   radiative decay $B_{s} \to \gamma \gamma$ and radiative leptonic decay $B_{s} \to \gamma \ell\bar{\ell}$ are of great importance to the determination of the parameters of the light-cone distribution amplitudes (LCDA) of $B_s$ meson, which is the fundamental nonperturbative input in the study on the $B_s$ meson decays. They are also sensitive to the new physics effect since they are induced by the flavor-changing-neutral current processes.  Relatively little attention has been paid to these modes due to their small branching ratios, while the current machines with high luminosity such as LHC and SuperKEKB have the capability to detect these decays \cite{Kou:2018nap}. The experimental progresses on these decay modes raise the necessity of more precisely theoretical studies, which have been improved in several aspects.

A comprehensive study on the $B_{d, \, s} \to \gamma \gamma$ decays is presented in \cite{Shen:2020hfq}, where both leading power contribution and power suppressed contributions to the decay amplitude have been analysed in detail. The leading power amplitude of $B_{d, \, s} \to \gamma \gamma$ can be factorized into the convolution of the effective Wilson coefficients, the jet function and the LCDA of $B$ meson \cite{Bosch:2002bv}, where both the effective Wilson coefficients and the jet function have been provided up to two-loop order in QCD \cite{Chetyrkin:1996vx, Liu:2020ydl}. The next-to-leading logarithm resummation has been performed within the framework of soft-collinear effective theory \cite{Bauer:2000yr,Beneke:2002ph}. Various subleading power contributions have also been investigated, including the power suppressed local contribution, the power suppressed nonlocal contribution from the hard-collinear propagator, the power suppressed term in the heavy quark expansion, the contribution from high twist LCDAs of $B$ meson, the strange quark mass effects and resolved photon contribution. Some power suppressed contributions are factorizable, such as the  nonlocal contributions from the hard-collinear propagator, from the heavy quark expansion, and  from  the high twist LCDAs, while the strange quark  mass contribution is not factorizable due to  endpoint singularity. 

As endpoint singularity appears in the convolution between the jet function and LCDA of $B$ meson in the contribution from the strange quark mass term in $B_s \to \gamma\gamma$ decay, the factorization approach cannot be applied. A parameterization method is employed in \cite{Shen:2020hfq}, which is model dependent with sizable theoretical uncertainty. In this paper we will adopt an alternative method with better predictive power. The basic idea of our method is to take the photon momentum (from the QED vertex) off-shell and replace the endpoint region of the convolution between the jet function and the LCDA of $B_s$ meson with part of the $B_s \to V$ ($V$ standing for a neutral vector meson) form factors relevant to the strange mass term, and  take the $q^2=0$ limit after the replacement. Since the correlation function should be expressed in terms of the dispersion integral in this method, it is usually called dispersion approach, which has been widely used in the evaluation of the power suppressed contribution in the exclusive processes such as $\gamma^{\ast} \gamma \to \pi$, $B \to \gamma\nu \ell$ et al. To estimate the contribution from the resolved  photon effect \cite{Khodjamirian:1997tk,Agaev:2010aq,Braun:2012kp,Wang:2016qii,Beneke:2018wjp,Shen:2020hsp}, and the result is consistent with the prediction from employing the photon LCDAs \cite{Ball:2002ps,Wang:2017ijn,Wang:2018wfj,Shen:2018abs,Shen:2019vdc,Shen:2019zvh}.

The $B_{d, \, s} \to \gamma \ell\bar{\ell}$ decays in the kinematic region of large photon energy is intensively analyzed with QCD factorization techniques recently \cite{Beneke:2020fot}, which constructs a systematic expansion in inverse powers of large photon energy and heavy quark mass. The next-to-leading power (NLP) corrections to the amplitudes  due to local and nonlocal A-type and B-type operators in the soft-collinear effective theory, as well as the local four-quark contributions, are also considered. The numerical calculation indicates that the NLP contribution can give rise to about $20\%-30\%$ correction to the branching ratios of the decay channels  \cite{Beneke:2020fot}.
The strange quark mass effect has not been investigated in the $B_{ s} \to \gamma \ell\bar{\ell}$ decays, and we will fill this gap with a similar method as that in the $B_s \to \gamma\gamma$ decay.

This article is organized as follows: In the next section we will review leading power and NLP contribution to the $B_{ s} \to \gamma \gamma$ and $B_{ s} \to \gamma \ell\bar{\ell}$ decays. In section 3 we will take advantage of the dispersion approach to evaluate the NLP contribution from quark mass term in the $B_{ s} \to \gamma \gamma$ and  $B_{ s} \to \gamma \ell\bar{\ell}$ decays. The numerical result is given in section 4. Summary is presented in the last section.


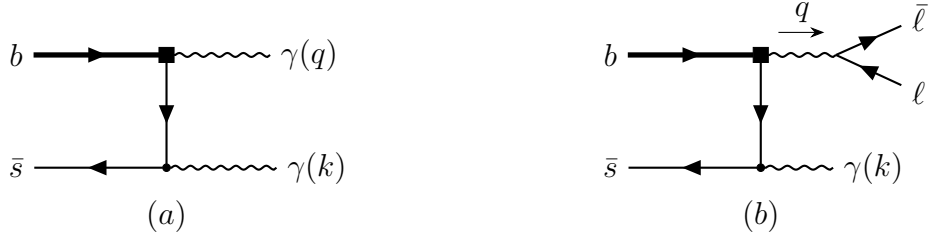
\begin{figure}
	\centering
\begin{tikzpicture}
\begin{feynman}[large]
\vertex (a) {\(b\)};
\vertex [square dot,scale=1, right=2cm of a] (b){};
\vertex [ right=1.9cm of b] (f1) {\(\gamma(q)\)};
\vertex [dot,scale=0.5,below=1.5cm of b] (c){};
\vertex [below=2.15cm of b] (cccc){\((a) \)};
\vertex [below=1.5cm of a] (f2) {\(\bar{s}\)};
\vertex [right=2cm of c] (f3) {\(\gamma(k)\)};
\diagram {
	(a) -- [fermion,line width=0.07cm] (b) -- [boson] (f1),
	(b) -- [fermion] (c),
	(c) -- [fermion] (f2),
	(c) -- [boson] (f3),
};
\end{feynman}
\begin{feynman}[large]
\vertex [right=4cm of f1](aa) {\(b\)};
\vertex [square dot,right=2cm of aa] (bb){};
\vertex [ right=1cm of bb] (ff1) ;
\vertex [dot,scale=0.5,below=1.5cm of bb] (cc){};
\vertex [below=2.15cm of bb] (ccc){\((b) \)};
\vertex [below=1.5cm of aa] (ff2) {\(\bar{s}\)};
\vertex [right=1.5cm of cc] (ff3) {\(\gamma(k)\)};
\vertex [right=2.1cm of cc] (ff4);
\vertex [above=0.7cm of ff4] (ff5){\(\ell\)};
\vertex [above=1cm of ff5] (ff6){\(\bar{\ell}\)};
\diagram {
	(aa) -- [fermion,line width=0.07cm] (bb) -- [boson,momentum=\(q \)] (ff1),
	(bb) -- [fermion] (cc),
	(cc) -- [fermion] (ff2),
	(ff5) -- [fermion ] (ff1),
	(ff1) -- [fermion] (ff6),
	(cc) -- [boson] (ff3),
};
\end{feynman}
\end{tikzpicture}
\caption{The leading order Feynman diagram of $\bar{B}_s\to \gamma\gamma$ (a) and $\bar{B}_s\to \gamma \ell\bar{\ell}$ (b) decay, where the other two diagrams due to the exchange of two gauge bosons are not presented.}\label{feynman}
\end{figure}

\section{The  amplitudes of  $\bar{B}_s\to \gamma\gamma$  and $\bar{B}_s\to \gamma\ell\bar{\ell}$ decays}
In order to express the decay amplitudes of the   $\bar{B}_s\to \gamma\gamma$  and $\bar{B}_s\to \gamma\ell\bar{\ell}$ decays, we start with the effective weak Hamiltonian where the unitarity of the CKM matrix has been employed
\begin{equation}
\label{hamiltonian1}
\mathcal{H}_{\mathrm{eff}}=\frac{4 G_{F}}{\sqrt{2}} \sum_{p=u, c} V_{p b} V_{p s}^{*}\left[\sum_{i=1}^{2} C_{i}(\nu) P_i^{(p)}(\nu)+\sum_{i=3}^{8} C_{i}(\nu) P_{i}(\nu)+{\alpha_{\rm em}\over 4\pi}\sum_{i=9}^{10} C_{i}(\nu) P_{i}(\nu)\right]+\text { h.c. }
\end{equation}
where the $P_{1,2}^{(p)}$ are four-quark tree operators, the $P_{3-6}$ are four quark QCD-penguin operators. The specific form of these operators and the corresponding Wilson coefficients $C_i$ can be found in \cite{Shen:2020hfq}. $P_7$ is the electro-magnetic penguin operator which leads to $b\to s\gamma$ transition at leading order in $\alpha_s$. $P_8$ is the chromo-magnetic penguin operator, and $P_{9,10}$ are semileptonic operators for $b\to q\ell\ell$ transitions. These four effective operators are listed as
\begin{equation}
\begin{aligned}
&P_{7}=-\frac{g_{\mathrm{em}} \bar{m}_{b}(\nu)}{16 \pi^{2}}\left(\bar{s}_{L} \sigma^{\mu \nu} b_{R}\right) F_{\mu \nu}, \quad P_{8}=-\frac{g_{s} \bar{m}_{b}(\nu)}{16 \pi^{2}}\left(\bar{s}_{L} \sigma^{\mu \nu} T^{a} b_{R}\right) G_{\mu \nu}^{a}\\
&P_{9}=-\left[\bar{s} \gamma^{\mu} P_{L} b\right]\left[\bar{\ell} \gamma_{\mu} \ell\right], \quad P_{10}=-\left[\bar{s} \gamma^{\mu} P_{L} b\right]\left[\bar{\ell} \gamma_{\mu} \gamma_{5} \ell\right],\quad P_L=(1-\gamma_5)/2,
\end{aligned}
\end{equation}
where $\bar{m}_{b}(\nu)$ is the $b$-quark mass in $\overline{\rm MS}$ scheme and the convention of the covariant derivative are the same with \cite{Shen:2020hfq}. The  amplitudes of  $\bar{B}_s\to \gamma\gamma$  and $\bar{B}_s\to \gamma\ell\bar{\ell}$ decays can be written by the matrix elements of the effective Hamiltonian, i.e.,
\begin{equation}
\begin{aligned}
\overline{\mathcal{A}}\left(\bar{B}_{s} \rightarrow \gamma \gamma\right)&=-\langle \gamma(k,\epsilon_1^\ast)\gamma(q,\epsilon_2^\ast)|\mathcal{H}_{\mathrm{eff}}|\bar B_s(k+q)\rangle,
 \\
\overline{\mathcal{A}}\left(\bar{B}_{s} \rightarrow \gamma \ell\bar{\ell}\right)&=-\langle \gamma(k,\epsilon^\ast)\ell(p_{\ell})\bar{\ell}(p_{ \bar\ell})|\mathcal{H}_{\mathrm{eff}}|\bar B_s(k+p_{\ell}+p_{\bar\ell})\rangle.
\label{matrix element}
\end{aligned}
\end{equation}

In the following, we quote the detailed expression of the decay amplitudes of $\bar{B}_s\to \gamma\gamma$  and $\bar{B}_s\to \gamma\ell\bar{\ell}$ processes which are obtained in \cite{Shen:2020hfq} and \cite{Beneke:2020fot} so that we can get the full amplitude after obtaining the contribution from quark mass term. For the $\bar{B}_s\to \gamma\gamma$ decays,  the  amplitude including the power suppressed contributions takes the form:
\begin{equation}
\mathcal{\overline{A}}\left(\bar{B}_{s} \rightarrow \gamma \gamma\right)=  -i \frac{ G_{F}\alpha_{\mathrm{em}}}{\sqrt{2} \pi} m_{B_{s}}^{3} \epsilon_{1}^{* \mu}(p) \epsilon_{2}^{* \nu}(q)\left[\left(g_{\mu \nu}^{\perp}-i \varepsilon_{\mu \nu}^{\perp}\right) \overline{\mathcal{A}}_{L}-\left(g_{\mu \nu}^{\perp}+i \varepsilon_{\mu \nu}^{\perp}\right) \overline{\mathcal{A}}_{R}\right],
\end{equation}
where  $\epsilon_{1}^{\star\mu}$ and $\epsilon_{2}^{\star\nu}$ stand for the polarization vector of the two outgoing photons. The shorthand notations $g_{\mu \nu}^{\perp} $ and $\epsilon_{\mu \nu}^{\perp} $ are defined as
\begin{eqnarray}
  g_{\mu \nu}^{\perp} \equiv g_{\mu \nu}-{n_{+\mu}   n_{-\nu} \over 2}
  -{n_{-\mu}   n_{+\nu} \over 2},  \quad
  \varepsilon_{\mu \nu}^{\perp} \equiv \frac{1}{2}\varepsilon_{\mu \nu \rho \tau}{ n_+^{\rho} n_-^{\tau}} = \varepsilon_{\mu \nu \rho \tau}  n_+^{\rho} v^{\tau}  \,,
  \quad
\end{eqnarray}
where the convention  $\varepsilon_{0123} = -1$ has been adopted. The light-cone vectors $n_+, n_-$ have been introduced which satisfy $n_+^2=0, n_-^2=0$ and $n_+\cdot  n_-=2$. The amplitudes $\bar {\cal A}_{L}$ and $\bar {\cal A}_{R}$ are classified according to the polarization of the final state photons. The manifest expressions of $\bar {\cal A}_{L}$ and $\bar {\cal A}_{R}$ can be derived in the following
\begin{eqnarray}
\bar {\cal A}_{L} &=& \sum_{p=u,  c} \, V_{p b} \, V_{p s}^{\ast} \,
\sum_{i=1}^8 \, C_i \, \left [ F_{i, \,  L}^{(p), \, \rm LP}
+  F_{i, \,  L}^{(p), \,  \rm  {fac, \, NLP}}
+  F_{i, \,  L}^{(p), \,  \rm  {soft, \, NLP}}   \right ] \,,
\nonumber \\
\bar {\cal A}_{R} &=& \sum_{p=u,  c} \, V_{p b} \, V_{p s}^{\ast} \,
\sum_{i=1}^8 \, C_i \, \left [ F_{i, \,  R}^{(p), \, \rm LP}
+  F_{i, \,  R}^{(p), \,  \rm  {fac, \, NLP}}
+  F_{i, \,  R}^{(p), \,  \rm  {soft, \, NLP}}   \right ] \,,
\label{definition: helicity amplitudes}
\end{eqnarray}
where the three terms in the square bracket denote the leading power contribution, the factorizable NLP contribution and the power suppressed soft contribution, respectively. The  factorizable NLP contribution collects various subleading power contributions, which is written as
\begin{eqnarray}
\sum_{i=1}^{8} \, C_i \, F_{i, \, L}^{(p), \,{\rm fac, \, NLP}}
&=& C_7^{\rm eff} \, \left [   F_{7, \, L}^{\rm hc, \, NLP}
+  F_{7, \, L}^{m_q, \, {\rm NLP}}
+  F_{7, \, L}^{A2, \, {\rm NLP}}
+ F_{7, \, L}^{\rm HT, \, NLP}
+   F_{7, \, L}^{e_b, \, {\rm NLP}}  \right ]
\nonumber \\
&& \, + \, {f_{B_s} \,  \over  m_{B_s}} \,
\left [ {\cal F}^{(p), \,{\rm WA}}_{V} - {\cal F}^{(p), \,{\rm WA}}_{A} \right ]  \,,
\nonumber \\
\sum_{i=1}^{8} \, C_i \, F_{i, \, R}^{(p), \,{\rm fac, \, NLP}}
&=& {f_{B_s} \,  \over  m_{B_s}} \,
\left [ {\cal F}^{(p), \,{\rm WA}}_{V} + {\cal F}^{(p), \,{\rm WA}}_{A} \right ]   \,.
\label{final form of the factorized NLP effect}
\end{eqnarray}
 From the above equation we can see that only the contribution from weak annihilation mechanism, which is induced by the four-quark operators can give rise to the amplitude with right-handed polarized photon, since the other contributions induced by electro-magnetic penguin operator are left-handed in nature. All the amplitudes in (\ref{final form of the factorized NLP effect}) have been derived in \cite{Shen:2020hfq} with great detail. For $\bar{B}_s \to \gamma\gamma$ decays, the  non-vanishing strange quark mass leads to the term $F_{7,L}^{m_s,\text{NLP}}$, which  will be specifically  investigated in the next section.

 For the $\bar{B}_{s} \to \gamma \ell\bar{\ell}$ decays, the decay amplitude  can be parameterized as
\begin{equation}
\begin{aligned}
\overline{\mathcal{A}}\left(\bar{B}_{s} \rightarrow \gamma \ell\bar{\ell}\right)=i e \frac{\alpha_{\mathrm{em}}G_F}{\sqrt{2} \pi} & E_{\gamma} \epsilon_{\mu}^{\star} \left[\left(g_{\perp}^{\mu \nu}-i \varepsilon_{\perp}^{\mu \nu}\right)\left(\overline{\mathcal{A}}_{L V}\left[\bar{u} \gamma_{\nu} v\right]+\overline{\mathcal{A}}_{L A}\left[\bar{u} \gamma_{\nu} \gamma_{5} v\right]\right)\right.\\
&\left.-\left(g_{\perp}^{\mu \nu}+i \varepsilon_{\perp}^{\mu \nu}\right)\left(\overline{\mathcal{A}}_{R V}\left[\bar{u} \gamma_{\nu} v\right]+\overline{\mathcal{A}}_{R A}\left[\bar{u} \gamma_{\nu} \gamma_{5} v\right]\right)\right],
\end{aligned}
\end{equation}
where $V$ and $A$ refer to the vector and axial-vector chirality structure
of the lepton currents, respectively. 
The helicity amplitudes are given by
\begin{align}
  \label{eq:F_LR^i}
  \overline{\cal A}_{h V} & = \sum_{p=u,  c}  V_{p b} \, V_{p s}^{\ast} \sum_{i=1}^9  C_i F_h^{(i)} ,
&
  \overline{\cal A}_{h A} & =\sum_{p=u,  c}  V_{p b} \, V_{p s}^{\ast}\,\, C_{10} F_h^{(10)} ,
&
  h & = L,R,
\end{align}
where the helicity form factors $F_h^{(i)}$ contain both the leading power contribution and NLP contributions with the $s\bar s$ resonance, which can be expressed by
\begin{equation}
F_h^{(i)}=F_h^{(i,\rm LP)}+F_h^{( i,\rm NLP)}+\mathcal{O}(\alpha_s^{2},\alpha_s\lambda^2,\lambda^4),  
\end{equation}
where $\lambda\equiv\Lambda_{\rm QCD}/E_{\gamma}$ and
the specific expression for these form factors have been given in \cite{Beneke:2020fot}. In this paper the non-vanishing quark mass contributions are denoted by $F_{m,h}^{(7A)}$, $F_{m,h}^{(7B)}$  and $F_{m,h}^{(9,10)}$ in order to distinguish the contribution with the photon emission from $P_7$ or $P_{9,10}$, and the subscript $h=L$ or $R$ according to the helicity.

\section{The contribution from the quark mass term with dispersion approach}
Before constructing the theoretical framework to evaluate the contribution from the quark mass term, we show explicitly the mass term in the hard-collinear quark propagator (see Fig. \ref{feynman}) as
  \begin{equation}
  \begin{aligned}
  \frac{\slashed{k}-\slashed{l}+m_s}{(k-l)^2-m_s^2+i0}&=
   \frac{\slashed{k}}{(k-l)^2+i0} \\ &+\frac{-\slashed{l}}{(k-l)^2+i0}+\frac{m_s}{(k-l)^2+i0}+\frac{m_s^2\slashed{k}}{[(k-l)^2+i0]^2} \\ &+\mathcal{O}\left(\frac{\Lambda_{QCD}^2}{m_{b}^{2}}\right) \label{quarkmassterm} .
  \end{aligned}
  \end{equation}
  According to our power counting rule, the soft momentum scales as $l\sim ({\lambda,\lambda,\lambda}$), and collinear momentum scales as $k\sim(\lambda^2,1,\lambda)$. As a result,  the leading power term, which is given in the first line of Eq. (\ref{quarkmassterm}) scales as $1/\lambda$. We assume the strange quark mass $m_s\sim \lambda$, then both the second and third term in the second line of Eq. (\ref{quarkmassterm}) are at subleading power, similar to the first term which has been extensively studied in \cite{Shen:2020hfq}. Note that the third term has been neglected in the previous study \cite{Shen:2020hfq}. A straightforward calculation of the second term at tree level leads to the following expression
\begin{eqnarray}
T_{7, \, \alpha \beta}^{m_s, \, {\rm NLP}} &=&
-\left [   i \, Q_s\, \overline{m}_b(\nu)  \, m_s \,  f_{B_s} \,  m_{B_s}  \right ] \,
\left [g_{\alpha \beta}^{\perp}  - i \, \varepsilon_{\alpha \beta}^{\perp} \right ] \,
\int_0^{\infty} \, d \omega \, {\phi_B^{-}(\omega, \mu)  \over \omega}  ,
\label{T7 NLP mq}
\end{eqnarray}
where the endpoint singularity appear, since the  LCDA $\phi_B^{-}(\omega, \mu)$ does not vanish when $\omega \to 0$. The endpoint singularity also emerges when the third term in the second line of Eq. (\ref{quarkmassterm}) is inserted to the integral in Eq.(\ref{T7 NLP mq}). In the following  we evaluate the mass term contribution to the  $\bar{B}_s\to \gamma\gamma$  and $\bar{B}_s\to \gamma\ell\bar{\ell}$ decays with dispersion approach.

\subsection{The contribution from quark mass term in $\bar{B}_s\to \gamma\gamma$}

We start from the correlation function
\begin{equation}
  \begin{aligned}
    \tilde T^{\mu\nu}_7(k,q)&=2 \bar{m}_{b}(\nu)\int d^{4} x e^{i k\cdot x}\left\langle 0\left|\mathrm{~T}\left\{j_{s}^{\nu}(x),\left[\bar{s}  \sigma^{\alpha \mu} q_{\alpha} P_{R} b\right](0)\right\}\right| \bar{B}_{s}(k+q)\right\rangle|_{k^2<0}\\
    &+[k \leftrightarrow q, \nu \leftrightarrow \mu],
   \end{aligned}
  \end{equation}
where $k$ is the momentum of the interpolation electro-magnetic current $j^{\mu}_s=Q_s[\bar{s}\gamma^{\mu}s]$. It is regarded to be a hard-collinear mode, i.e. $|k^2|\sim m_b\Lambda$ and $k^2<0$,  so that the correlation function can be calculated using perturbation approach. The mass dependent term of the correlation function is totally left-handed, therefore it takes the following form
  \begin{equation}
  \begin{aligned}
   \tilde{T}_{7,\text{NLP}}^{\mu \nu,m_s}=i(g^{\mu\nu}_{\perp}-i\epsilon^{\mu\nu}_{\perp}) \tilde{F}^{m_s}_{7,\text{NLP}} ,
   \end{aligned}
  \end{equation}
where the scalar correlation function can be expressed as the factorization form after a calculation on the partonic level
   \begin{equation}
  \begin{aligned}
   \tilde{F}^{m_s}_{7,\text{NLP}}
  =-\left(m_s\bar{m}_{b}Q_s f_{B_s}m_{B_s}\frac{n_-q}{n_+k}\right)\left[\int_{0}^{\infty}\frac{\phi_{B}^-(\omega)}{\omega-n_-k}d\omega
  -m_s\int_{0}^{\infty}\frac{\phi_{B}^+(\omega)}{(\omega-n_-k)^2}d\omega\right].
  \end{aligned}\label{pl1}
  \end{equation}
The $n_-k$ in the denominator regularize the endpoint singularity, at the cost of the nonphysical offshellness of the photon. The on-shell limit is to be taken after removing the singularity, then the on-shell condition will be recovered and the correlation function turns to the physical matrix element. The correlation function will be also expressed in terms of hadronic parameters after inserting the complete set of hadronic states and isolate the ground state contribution. The hadronic form factors relevant to the $B \to V$ transitions induced by the tensor current are defined as

  \begin{equation}
  \begin{aligned}
  \left\langle V\left(k, \epsilon^{*}\right)\left|\bar{q} \sigma_{\mu \nu} q^{\nu} b\right| \bar{B}(k+q)\right\rangle=& 2 a_{V}^{(q)} T_{1}\left(q^{2}\right) \varepsilon_{\mu \nu \rho \sigma} \epsilon^{* \nu}(k) k^{\rho} q^{\sigma}, \\
  \left\langle V\left(k, \epsilon^{*}\right)\left|\bar{q} i \sigma_{\mu \nu} \gamma_{5} q^{\nu} b\right| \bar{B}(k+q)\right\rangle=& a_{V}^{(q)} T_{2}\left(q^{2}\right)\left[\left(m_{B}^{2}-m_{V}^{2}\right) \epsilon_{\mu}^{*}(k)-\left(\epsilon^{*} \cdot q\right)(2 k+q)_{\mu}\right] \\
+&a_{V}^{(q)} T_{3}\left(q^{2}\right)\left(\epsilon^{*} \cdot q\right)\left[q_{\mu}-\frac{q^{2}}{m_{B}^{2}-m_{V}^{2}}(2 k+q)_{\mu}\right],
  \end{aligned}
  \end{equation}
  where the flavour factor $a_V^{(q)}$ comes from the quark structure of the vector mesons, and in this work $a_{\phi}^{(s)}=1$ will be employed.  The hadronic representation of the correlation functions then reads:
  \begin{equation}
  \begin{aligned}
    \tilde T^{\mu \nu}_7
  &=i(g^{\mu\nu}_{\perp}-i\epsilon^{\mu\nu}_{\perp})\left\{\frac{Q_s f_V m_Vm_{B_s}\bar{m}_{b}}{m_V^2-k^2}[n_+kT_1(q^2)+m_{B_s}T_2(q^2)]
  +{1\over \pi}\int_{\omega_s}^{\infty} d \omega^{\prime} \,
{\rho^{\rm had}(\omega^{\prime}) \over \omega^{\prime} - n_-k - i 0}\right\}.
\label{cfha1}\end{aligned}
\end{equation}
In the present work, we are only interested in the strange quark mass dependent part of the correlation function, then the scalar correlation function $\tilde{F}^{m_s}_{\text{NLP}}$ at the hadronic level is then written by
\begin{equation}
  \begin{aligned}
   \tilde{F}^{m_s}_{7,\text{NLP}}
  &=\frac{Q_s f_V m_Vm_{B_s}\bar{m}_{b}}{m_V^2-k^2}[n_+kT^s_1(q^2)+m_{B_s}T^s_2(q^2)]
  +{1\over \pi}\int_{\omega_s}^{\infty} d \omega^{\prime} \,
{\rho^s_{\rm had}(\omega^{\prime}) \over \omega^{\prime} - n_-k - i 0}.
\label{cfha}\end{aligned}
\end{equation}
  Matching the two different representations of the correlation functions with the parton-hadron duality, namely, equalizing the dispersion integral in the QCD expression and the hadronic expression of the correlation function above the threshold, and performing the Borel transformation with respect to the variable $n_-k$, we obtain the sum rules for the form factors relative to the strange quark mass term
  	\begin{equation}
  \begin{aligned}
  n_+kT_1^{(m_s)}(q^2)+m_{B_s}T_2^{(m_s)}(q^2)&=-\frac{m_sf_{B_s}n_-q}{f_Vm_V}\int_0^{\omega_{s}} \exp \left(\frac{m_{V}^{2}-n_+k \omega^{\prime}}{n_+k \omega_{M}}\right)\phi_B^-( \omega^{\prime})d\omega^{\prime} \\
  &-\frac{m_s^2f_{B_s}n_-q}{f_Vm_V}\int_0^{\omega_{s}} \exp \left(\frac{m_{V}^{2}-n_+k \omega^{\prime}}{n_+k \omega_{M}}\right)\frac{\partial \phi_{B}^{+}\left(\omega^{\prime}\right)}{\partial \omega^{\prime}}d\omega^{\prime}.
  \end{aligned}\label{ff1}
  \end{equation}
The procedure of evaluation of the above form factors is actually the $B$-meson light-cone sum rules\cite{Khodjamirian:2005ea,DeFazio:2005dx}, which has been used to calculate various heavy-to-light transition form factors\cite{Wang:2015vgv,Shen:2016hyv,Wang:2017jow,Lu:2018cfc,Gao:2019lta,Wang:2015ndk}. After obtaining the specific expression of  the term  $n_+kT_1^{(m_s)}(q^2)+m_{B_s}T_2^{(m_s)}(q^2)$, we replace the continuum states contribution by the QCD result above the threshold in (\ref{cfha}).    Taking the limit $n_-k\to 0$,   the scalar correlation function $\tilde{F}^{m_s}_{7,\text{NLP}}$ turns to the physical amplitude $F_{7, L}^{m_s, \mathrm{NLP}}$
\begin{equation}
\begin{aligned}
{F}^{m_s,\text{NLP}}_{7,L}=\frac{Q_s f_V m_{B_s}\bar{m}_{b}}{m_V}[n_+kT_1^{(m_s)}(q^2)+m_{B_s}T_2^{(m_s)}(q^2)]+\int_{\omega_{s}}^{\infty}\frac{\operatorname{Im} \tilde{F}_{7,\text{NLP}}^{m_s}}{\omega^{\prime}}d\omega^{\prime}.
\label{ma1}\end{aligned}
\end{equation}
Substituting Eq. (\ref{ff1}) and  Eq. (\ref{pl1}) with $n_-k \to 0$ into Eq. (\ref{ma1}), we arrive at the final expression of the strange mass term  with dispersion approach as  
  \begin{equation}
  \begin{aligned}
F_{7, L}^{m_s, \mathrm{NLP}} &=-\frac{Q_{s} f_{B_{s}} \bar{m}_{b} m_{s}}{ m_{B_{s}}^{2}}\left\{\frac{m_{B_{s}}}{m_{\phi}^{2}} \int_{0}^{\omega_{s}} \exp \left(\frac{m_{\phi}^{2}-m_{B_{s}} \omega^{\prime}}{m_{B_{s}} \omega_{M}}\right) \phi_{B}^{-}\left(\omega^{\prime}\right) d \omega^{\prime}+\int_{\omega_{s}}^{\infty} \frac{\phi_{B}^{-}\left(\omega^{\prime}\right)}{\omega^{\prime}} d \omega^{\prime}\right. \\
  &- \left.\frac{m_{B_{s}}}{m_{\phi}^{2}} \int_{0}^{\omega_{s}} \exp \left(\frac{m_{\phi}^{2}-m_{B_{s}} \omega^{\prime}}{m_{B_{s}} \omega_{M}}\right) \frac{\partial \phi_{B}^{+}\left(\omega^{\prime}\right)}{\partial \omega^{\prime}} d \omega^{\prime}-\int_{\omega_{s}}^{\infty} \frac{1}{\omega^{\prime}} \frac{\partial \phi_{B}^{+}\left(\omega^{\prime}\right)}{\partial \omega^{\prime}} d \omega^{\prime}\right\}, \\
   \end{aligned}
  \end{equation}
  where we   used $n_+k=n_-q=m_{B_s}$ and $m_V\equiv m_{\phi}$, and the threshold parameter and Borel parameter
  \begin{equation}
  \omega_{s}=\frac{s_{0}}{n_+k} \sim \mathcal{O}\left(\frac{\Lambda_{\mathrm{QCD}}^{2}}{m_{B_{s}}}\right), \quad \omega_{M}=\frac{M^{2}}{n_+k} \sim \mathcal{O}\left(\frac{\Lambda_{\mathrm{QCD}}^{2}}{m_{B_{s}}}\right).
  \end{equation}
  Noted here, the well-known hierarchy structure of weak interaction also emerges in quark mass corrections clearly.

\subsection{The contribution of quark mass term in $\bar{B}_s\to \gamma \ell\bar{\ell}$}

 The subleading power contribution to the $\bar{B}_s\to \gamma \ell\bar{\ell}$ decays from strange quark mass effect is not considered in \cite{Beneke:2020fot}, therefore, we formally work out the factorization formula first. Following \cite{Beneke:2020fot}, we divide the correlation functions relevant to this process into A-type and B-type, which take the following form respectively
\begin{equation}
\begin{aligned}
T_{i}^{\mu \nu} &=\int d^{4} x e^{i k x}\left\langle 0\left|\mathrm{~T}\left\{Q_{s}\left[\bar{s} \gamma^{\nu} s\right](x),\left[\bar{q} \gamma^{\mu} P_{L} b\right](0)\right\}\right| \bar{B}_{s}\right\rangle, \quad i=9,10 \\
T_{7 A}^{\mu \nu} &=\frac{2 \bar{m}_{b}}{q^{2}} \int d^{4} x e^{i k x}\left\langle 0\left|\mathrm{~T}\left\{Q_{s}\left[\bar{s} \gamma^{\nu} s\right](x),\left[\bar{q} i \sigma^{\mu \alpha} q_{\alpha} P_{R} b\right](0)\right\}\right| \bar{B}_{s}\right\rangle, \\
T_{7 B}^{\mu \nu} &=\frac{2 \bar{m}_{b}}{q^{2}} \int d^{4} x e^{i q x}\left\langle 0\left|\mathrm{~T}\left\{Q_{s}\left[\bar{s} \gamma^{\mu} s\right](x),\left[\bar{q} i \sigma^{\nu \alpha} k_{\alpha} P_{R} b\right](0)\right\}\right| \bar{B}_{s}\right\rangle.
\end{aligned}
\end{equation}
After contraction of the strange quark field, it is easy to isolate the quark mass term and obtain the related factorization formula directly 
\begin{equation}
\begin{aligned}
\tilde{T}_{9,10}^{\mu \nu,m_s} &=\left(g_{\perp}^{\mu \nu}-i \epsilon_{\perp}^{\mu \nu}\right)\frac{Q_{s} f_{B_{s}} m_{B_{s}}}{4} \frac{m_{s}}{n_{+} k} \int_{0}^{\infty}\left[ \frac{\phi_{B}^{-}(\omega)}{\omega}-m_s\frac{\phi_{B}^{+}(\omega)}{\omega^{2}}\right] d \omega \\
&+\left(g_{\perp}^{\mu \nu}+i \epsilon_{\perp}^{\mu \nu}\right) \frac{Q_{s} f_{B_{s}} m_{B_{s}}}{4} \frac{m_{s}}{n_+k} \int_{0}^{\infty} \frac{\phi_{B}^{+}(\omega)}{\omega} d \omega, \\
\tilde{T}_{7A}^{\mu \nu,m_s}
&=-\left(g_{\perp}^{\mu \nu}-i \epsilon_{\perp}^{\mu \nu}\right) \frac{2 \bar{m}_{b}}{q^{2}} \frac{Q_{s} f_{B_{s}} m_{B_{s}}n_-q}{4} \frac{m_{s}}{n_+k} \int_{0}^{\infty} \left[ \frac{\phi_{B}^{-}(\omega)}{\omega}-m_s\frac{\phi_{B}^{+}(\omega)}{\omega^{2}}\right] d \omega \\
&-\left(g_{\perp}^{\mu \nu}+i \epsilon_{\perp}^{\mu \nu}\right) \frac{2 \bar{m}_{b}}{q^{2}} \frac{Q_{s} f_{B_{s}} m_{B_{s}}n_+q}{4} \frac{m_{s}}{n_+k} \int_{0}^{\infty} \frac{\phi_{B}^{+}(\omega)}{\omega} d \omega, \\
\tilde{T}_{7B}^{\mu \nu,m_s} &=-\left(g_{\perp}^{\mu \nu}-i \epsilon_{\perp}^{\mu \nu}\right) \frac{2 \bar{m}_{b}}{q^{2}} \frac{Q_{s} f_{B_{s}} n_+k m_{s}}{4} \int_{0}^{\infty}\left[ \frac{\phi_{B}^{-}(\omega)}{\omega-n_+q}-\frac{m_s\phi_{B}^{+}(\omega)}{\left(\omega-n_+q\right)^{2}}\right] d \omega.
\end{aligned}
\end{equation}
 From the above equations, it is obvious that there is no endpoint singularity in B-type contribution for the existence of $n_+q$ in the denominator, and no endpoint singularity in the right-handed amplitudes since they are proportional to the first inverse moment $1/\lambda_{B_s}$. While for the other amplitudes, the endpoint singularity arises due to the endpoint behavior of the LCDAs of $B$-meson.  Following the same procedure with $\bar{B}_s \to \gamma\gamma$ decay, we start from the correlation function in which the momentum related to the electromagnetic current is taken to be off mass shell, then the endpoint singularity  is regularized. For the convolution with the integral variable below the threshold parameter,  we can express the correlation functions with the help of hadronic form factors
\begin{equation}
\begin{aligned}
T_{7A}^{\mu \nu,s}|_{\omega<\omega_s}&=\frac{1}{2}\left(g_{\perp}^{\mu \nu}-i \epsilon_{\perp}^{\mu \nu}\right)\frac{\bar{m}_{b}}{q^{2}} \frac{Q_{s} f_{V} m_{V}m_{B_s}}{m_{V}^{2}-k^{2}} \left[n_+k  T^{(m_s)}_{1}\left(q^{2}\right)+m_{B_{s}} T^{(m_s)}_{2}\left(q^{2}\right)\right],\\
T_{9,10}^{\mu \nu,s}|_{\omega<\omega_s}&=\frac{1}{4} \left(g^{\mu \nu}-i \epsilon_{\perp}^{\mu \nu}\right)\frac{f_{V} m_{V} Q_{s}}{m_{V}^{2}-k^{2}}\left[n_+k V^{(m_s)}\left(q^{2}\right)+m_{B_{s}} A^{(m_s)}_{1}\left(q^{2}\right)\right],
\end{aligned}
\end{equation}
where the $B \to V$ form factors induced by the vector and axial vector current are defined by
\begin{equation}
  \begin{aligned}
   \left\langle V\left(k, \epsilon^{*}\right)\left|\bar{q} \gamma_{\mu} b\right| \bar{B}(k+q)\right\rangle=&-\frac{2 ia_V^{(q)} V\left(q^{2}\right)}{m_{B}+m_{V}} \epsilon_{\mu \nu \rho \sigma} \epsilon^{* \nu} k^{\rho} q^{\sigma} ,\\
  \left\langle V\left(k, \epsilon^{*}\right)\left|\bar{q} \gamma_{\mu} \gamma_{5} b\right| \bar{B}(k+q)\right\rangle&= \frac{2 m_{V} \epsilon^{*} \cdot q}{q^{2}} q_{\mu} a_V^{(q)}A_{0}\left(q^{2}\right) \\
  &+\left(m_{B}+m_{V}\right)\left[\epsilon_{\mu}^{*}-\frac{\epsilon^{*} \cdot q}{q^{2}} q_{\mu}\right] a_V^{(q)}A_{1}\left(q^{2}\right)\\
  &-\frac{\epsilon^{*} \cdot q}{m_{B}+m_{V}}\left[(2 k+q)_{\mu}-\frac{m_{B}^{2}-m_{V}^{2}}{q^{2}} q_{\mu}\right] a_V^{(q)}A_{2}\left(q^{2}\right). \\
    \end{aligned}
  \end{equation}
 The form factors related to the strange quark mass can be obtained using the same method with the previous subsection. After   making the Lorentz decomposition
\begin{equation}
T_{i}^{\mu \nu}(k, q)=E_{\gamma}\left[g_{\perp}^{\mu \nu}\left(F_{L}^{(i)}-F_{R}^{(i)}\right)-i \varepsilon_{\perp}^{\mu \nu}\left(F_{L}^{(i)}+F_{R}^{(i)}\right)\right],
\end{equation}
the final factorized expressions for the quark mass corrections to the quark mass dependent amplitudes $F_{m,L}^{i}$ are then obtained as
\begin{equation}
\begin{aligned}
F_{m,L}^{(9,10)}&=\frac{Q_s f_{B_s}m_{B_s}m_s}{8E_{\gamma}^2}  \left\{ \int_{0}^{\omega_{s}} d \omega^{\prime} \frac{2E_{\gamma}}{m_{\phi}^{2}} \exp \left(\frac{m_{\phi}^{2}-2E_{\gamma} \omega^{\prime}}{2E_{\gamma} \omega_{M}}\right)\phi_B^-( \omega^{\prime}) +\int_{\omega_{s}}^{\infty} \frac{\phi_B^-( \omega^{\prime})}{\omega^{\prime}}d \omega^{\prime}
\right\}
\\
+ &\frac{Q_s f_{B_s}m_{B_s}m_s^2}{8E_{\gamma}^2}\left\{\int_{0}^{\omega_{s}} d \omega^{\prime} \frac{2E_{\gamma}}{m_{\phi}^{2}} \exp \left(\frac{m_{\phi}^{2}-2E_{\gamma} \omega^{\prime}}{2E_{\gamma} \omega_{M}}\right)\frac{\partial\phi_B^+(\omega')}{\partial\omega'}+\int_{\omega_{s}}^{\infty} \frac{1}{\omega^{\prime}}\frac{\partial\phi_B^+(\omega)}{\partial\omega}d \omega^{\prime}
\right\},\\
F_{m,L}^{(7A)}&=-\frac{2 \bar{m}_{b}}{q^{2}}\frac{Q_s f_{B_s}m_{B_s}^2m_s}{8E^2_{\gamma}} \left\{ \int_{0}^{\omega_{s}} d \omega^{\prime} \frac{2E_{\gamma}}{m_{\phi}^{2}} \exp \left(\frac{m_{\phi}^{2}-2E_{\gamma} \omega^{\prime}}{2E_{\gamma} \omega_{M}}\right)\phi_B^-( \omega^{\prime})+\int_{\omega_{s}}^{\infty} \frac{\phi_B^-( \omega^{\prime})}{\omega^{\prime}}d \omega^{\prime}
\right\}\\
- &\frac{2 \bar{m}_{b}}{q^{2}}\frac{Q_s f_{B_s}m_{B_s}^2m_s^2}{8E^2_{\gamma}} \left\{ \int_{0}^{\omega_{s}} d \omega^{\prime} \frac{2E_{\gamma}}{m_{\phi}^{2}} \exp \left(\frac{m_{\phi}^{2}-2E_{\gamma} \omega^{\prime}}{2E_{\gamma} \omega_{M}}\right)\frac{\partial\phi_B^+(\omega')}{\partial\omega'}+\int_{\omega_{s}}^{\infty} \frac{1}{\omega^{\prime}}\frac{\partial\phi_B^+(\omega)}{\partial\omega}d \omega^{\prime}
\right\},\\
F_{m,L}^{(7B)}&=-\frac{2 \bar{m}_{b}}{q^{2}} \frac{Q_{s} f_{B_{s}} m_{s}}{2} \int_{0}^{\infty} \frac{\phi_{B}^{-}(\omega)}{\omega-q^{2} / m_{B_{s}}} d \omega-\frac{2 \bar{m}_{b}}{q^{2}} \frac{Q_{s} f_{B_{s}} m_{s}^{2}}{2}\\
&\left\{ \int_{0}^{\omega_{s}} d \omega^{\prime} \frac{m_{B_s}}{m_{\phi}^{2}-q^2} \exp \left(\frac{m_{\phi}^{2}-m_{B_s} \omega^{\prime}}{m_{B_s} \omega_{M}}\right)\frac{\partial\phi_B^+(\omega')}{\partial\omega'}+\int_{\omega_{s}}^{\infty} \frac{1}{\omega^{\prime}-q^2/m_{B_s}}\frac{\partial\phi_B^+(\omega)}{\partial\omega}d \omega^{\prime}
\right\},\\
\end{aligned}
\end{equation}
where the second term of $F_{m,L}^{(7B)}$ represents the contribution of $\phi(s\bar{s})$ resonance and can be incorporated into $\overline{\mathcal{A}}_{L V}^{(7 B)}\left(q^{2}\right)$ in \cite{Beneke:2020fot}. So this term has been omitted in the following numerical analysis.

The other amplitudes are factorizable, so that we present the factorization formula as follows
\begin{equation}
\begin{aligned}
F_{m,R}^{(9,10)}&=\frac{Q_{s} f_{B_{s}} m_{B_{s}}m_s}{8E_{\gamma}^2}  \int_{0}^{\infty} \frac{\phi_{B}^{+}(\omega)}{\omega} d \omega ,\\
F_{m,R}^{(7A)}&=-\frac{2 \bar{m}_{b}}{q^{2}} \frac{Q_{s} f_{B_{s}} m_{B_{s}}\left(m_{B_{s}}-2 E_{\gamma}\right)m_s}{8E_{\gamma}^2}  \int_{0}^{\infty} \frac{\phi_{B}^{+}(\omega)}{\omega} d \omega ,\\
F_{m,R}^{7B}&=0.
\end{aligned}
\end{equation}
The helicity amplitudes are given by (\ref{eq:F_LR^i}) with
\begin{equation}
\sum_{i=1}^{9}\eta_{i} C_{i} F_{m,h}^{(i)}=C_7^{\text{eff}}(F_{m.h}^{(7A)}+F_{m.h}^{(7B)})+C_9^{\text{eff}}F_{m.h}^{(9)} \qquad h=L,R.
\end{equation}

\section{Numerical analysis}

\begin{table}[h]
\begin{center}
\caption{Input parameters in the numerical calculations.}
\label{parameters}
\renewcommand\arraystretch{1.8}
\begin{tabular}{|l|lr||l|lr|}
\hline
\text{Parameter}&\text{Value}&Ref.&Parameter&Value&\text{Ref.}\\
\hline
\hline
$m_{B_s}$ & 5.36688 GeV & \cite{ParticleDataGroup:2020ssz} &$m_{\phi}$ & 1.01946 GeV & \cite{ParticleDataGroup:2020ssz}\\
$f_{B_s}|_{N_f=2+1+1}$&230.3 MeV&\cite{FlavourLatticeAveragingGroup:2019iem}&$\bar{m}_b$(4.8 GeV)&4.101 GeV&\cite{ParticleDataGroup:2020ssz}\\
$\bar{m}_s$(2 GeV)&92.9$\pm$0.7 MeV&\cite{ParticleDataGroup:2020ssz}&$\lambda_{B_s}$&0.40$\pm$0.15 GeV&     $\cite{Beneke:2020fot}$\\
$\alpha_s^{(5)}(m_Z)$&0.1188$\pm$0.0017 & \cite{ParticleDataGroup:2020ssz} & $\tau_{B_s}$ & (1.527$\pm$0.011)ps &\cite{ParticleDataGroup:2020ssz}\\
\hline
\{$\hat{\sigma}^{(1)}_{B_s}(\mu_0)$,$\hat{\sigma}^{(2)}_{B_s}(\mu_0)$\} & \{0.0$\pm$0.7,0.0$\pm$6.0\} & $\cite{Beneke:2020fot}$ & \{$M^2_{\phi}$,$s^0_{\phi}$\} & \{1.9$\pm$0.5,1.6$\pm$0.1\} GeV$^2$ &$\cite{Gao:2019lta}$
\\
\hline
\end{tabular}
\end{center}
\end{table}

In this section we will evaluate the numerical results of the quark mass contribution to the $\bar{B}_s\to \gamma\gamma$  and $\bar{B}_s\to \gamma\ell\bar{\ell}$ decays.   We firstly discuss the nonperturbative hadronic inputs entering the factorized expressions
of the helicity amplitudes. The leptonic decay constant of the  $B_s$ meson is taken from the averages values of Lattice simulation  \cite{Bazavov:2017lyh}. The two-particle $B_s$ meson distribution amplitudes in HQET serve as the fundamental ingredients   in the factorization formulae and the expression of the amplitudes from the dispersion approach.
Following ref.\cite{Beneke:2018wjp} we will introduce the general three-parameter ansatz for
the leading-twist LCDA  $\phi_B^{+}(\omega, \mu_0)$
\begin{eqnarray}
\phi_B^{+}(\omega, \mu_0) &=& \int_0^{\infty} d s \, \sqrt{w \, s} \,\,
J_{1}(2 \, \sqrt{w \, s}) \, \eta_{+}(s, \mu_0) \,,
\nonumber \\
\eta_{+}(s, \mu_0) &=&  {}_{1}F_{1}(\alpha; \beta; -s \, \omega_0) \,,
\label{model of B-meson DA}
\end{eqnarray}
where $J_1$ is the Bessel function, and ${}_{1}F_{1}$ is a   hypergeometric function.
It is useful to define the first inverse moment  and the inverse-logarithmic moments of the leading-twist $B_s$-meson LCDA
\begin{eqnarray}
\lambda_{B_s}^{-1}(\mu)& = &  \int_0^{\infty} \, d \omega \,\,
{\phi_B^{+}(\omega, \mu)  \over \omega}\nonumber \\
\widehat{\sigma}_{B_s}^{(n)} (\mu) &=& \lambda_{B_s}(\mu) \,
\int_{0}^{\infty} \, {d \omega \over \omega} \,
\left [ \ln \left ( {\lambda_{B_s}(\mu) \over \omega} \right )
- \gamma_E \right ]^{n} \, \phi_{B}^{+}(\omega, \mu)  \,.
\label{definition：inverse-logarithmic moment}
\end{eqnarray}
The shape parameter $\lambda_{B_s}$ and the associated inverse-logarithmic moments $\widehat{\sigma}_{B_s}^{(n)}$
are related to the parameters $\omega_0,\alpha,\beta$ with the following identities
\begin{eqnarray}
\lambda_{B_s}(\mu_0) &=& \left ( {\alpha-1 \over \beta-1} \right ) \, \omega_0 \,,
\qquad
\widehat{\sigma}_{B_s}^{(1)} (\mu_0) =  \psi (\beta-1) - \psi (\alpha-1)
+ \ln  \left ( {\alpha-1 \over \beta-1} \right ) \,,
\nonumber \\
\widehat{\sigma}_{B_s}^{(2)} (\mu_0) &=&
\left [  \widehat{\sigma}_{B_s}^{(1)} (\mu_0)  \right ]^2 +
\psi^{(1)}(\alpha-1) - \psi^{(1)}(\beta-1)  + {\pi^2 \over 6} \,,
\end{eqnarray}
so that the parameter $\alpha,\beta,\omega_0$ can be determined through $\lambda_{B_s}$ and  $\widehat{\sigma}_{B_s}^{(n)}$. However, even the inverse moment $\lambda_{B_s}$   has not been satisfactorily constrained albeit that distinct techniques and strategies
has been employed to investigate this parameter since it is defined by a nonlocal operator\cite{Wang:2016qii,Beneke:2018wjp,Wang:2018wfj,Wang:2015vgv,Wang:2017jow,Braun:2003wx,Li:2012nk,Li:2012md,Beneke:2011nf}.
Consequently, we will vary the input value of $\lambda_{B_s}$ in the conservative interval
as presented in Table \ref{parameters}. For the  associated inverse-logarithmic moments $\widehat{\sigma}_{B_s}^{(n)}$, we adopt the same values as \cite{Shen:2020hfq}. The two-particle twist-3 LCDA of $B$-meson is also of great importance in our calculation, we adopt the following model
\begin{eqnarray}	
\phi_{B}^{-\mathrm{WW}}\left(\omega, \mu_{0}\right) &=&\int_{\omega}^{\infty} d \rho f(\rho) \\
\phi_{B}^{-\mathrm{tw} 3}\left(\omega, \mu_{0}\right) &=&\frac{1}{6} \kappa\left(\mu_{0}\right)\left[\lambda_{E}^{2}\left(\mu_{0}\right)-\lambda_{H}^{2}\left(\mu_{0}\right)\right]\left[\omega^{2} f^{\prime}(\omega)+4 \omega f(\omega)-2 \int_{\omega}^{\infty} d \rho f(\rho)\right]
\end{eqnarray}
with
\begin{equation}
\begin{aligned}
\int_{0}^{\infty} d \omega f(\omega) &=\lambda_{B_{q}}^{-1}\left(\mu_{0}\right), \quad \int_{0}^{\infty} d \omega \omega f(\omega)=1, \quad \int_{0}^{\infty} d \omega \omega^{2} f(\omega)=\frac{4}{3} \bar{\Lambda} \\
\kappa^{-1}\left(\mu_{0}\right) &=\frac{1}{2} \int_{0}^{\infty} d \omega \omega^{3} f(\omega)=\bar{\Lambda}^{2}+\frac{1}{6}\left[2 \lambda_{E}^{2}\left(\mu_{0}\right)+\lambda_{H}^{2}\left(\mu_{0}\right)\right].
\end{aligned}
\end{equation}
For the other parameters such as the running quark mass, the threshold parameter and the Borel mass, we also follow the same choice as \cite{Shen:2020hfq}. The specific vales are given in Table \ref{parameters}. 

\begin{figure}
\begin{center}
\includegraphics[width=0.45 \columnwidth]{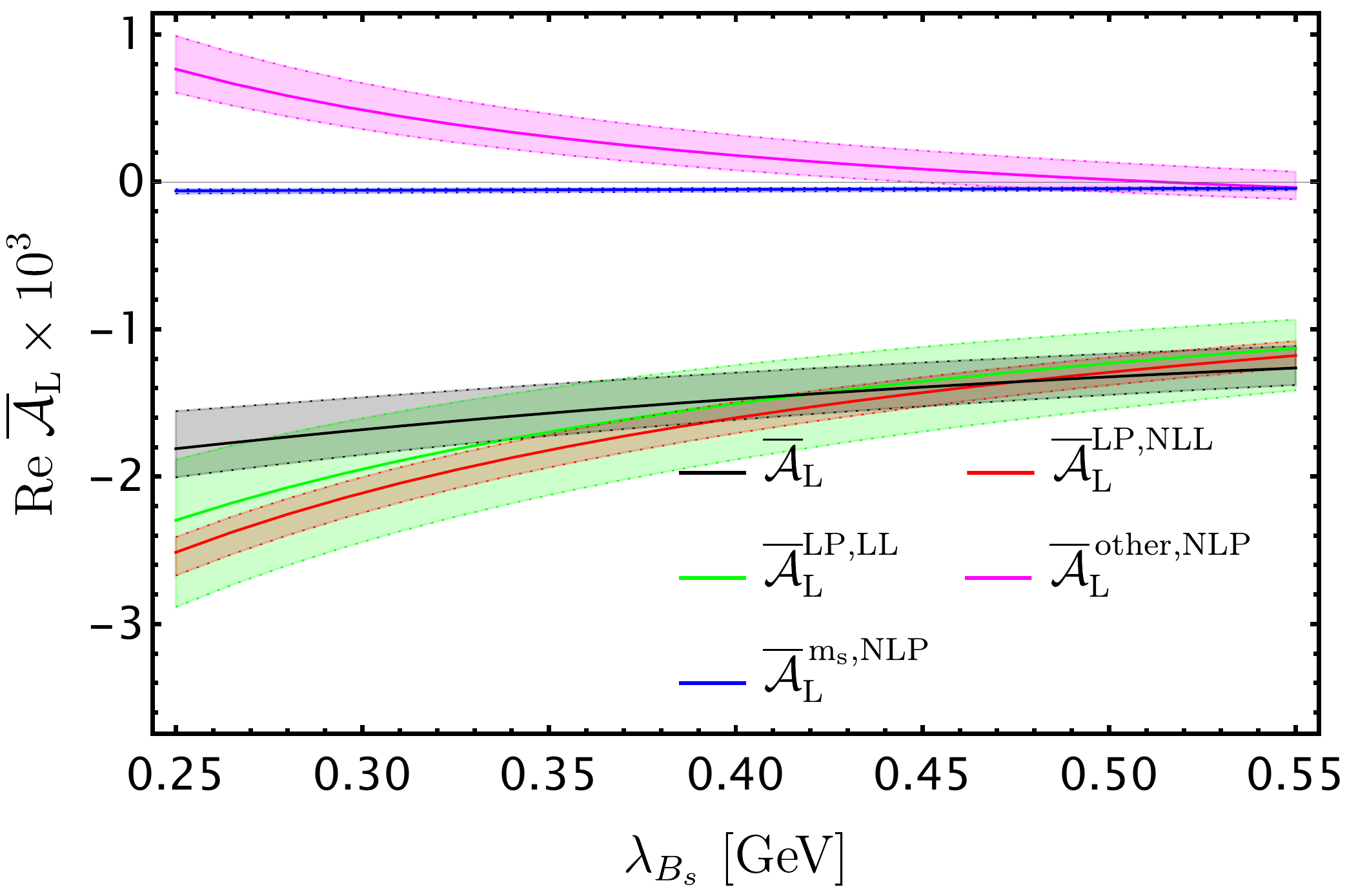}
\includegraphics[width=0.47 \columnwidth]{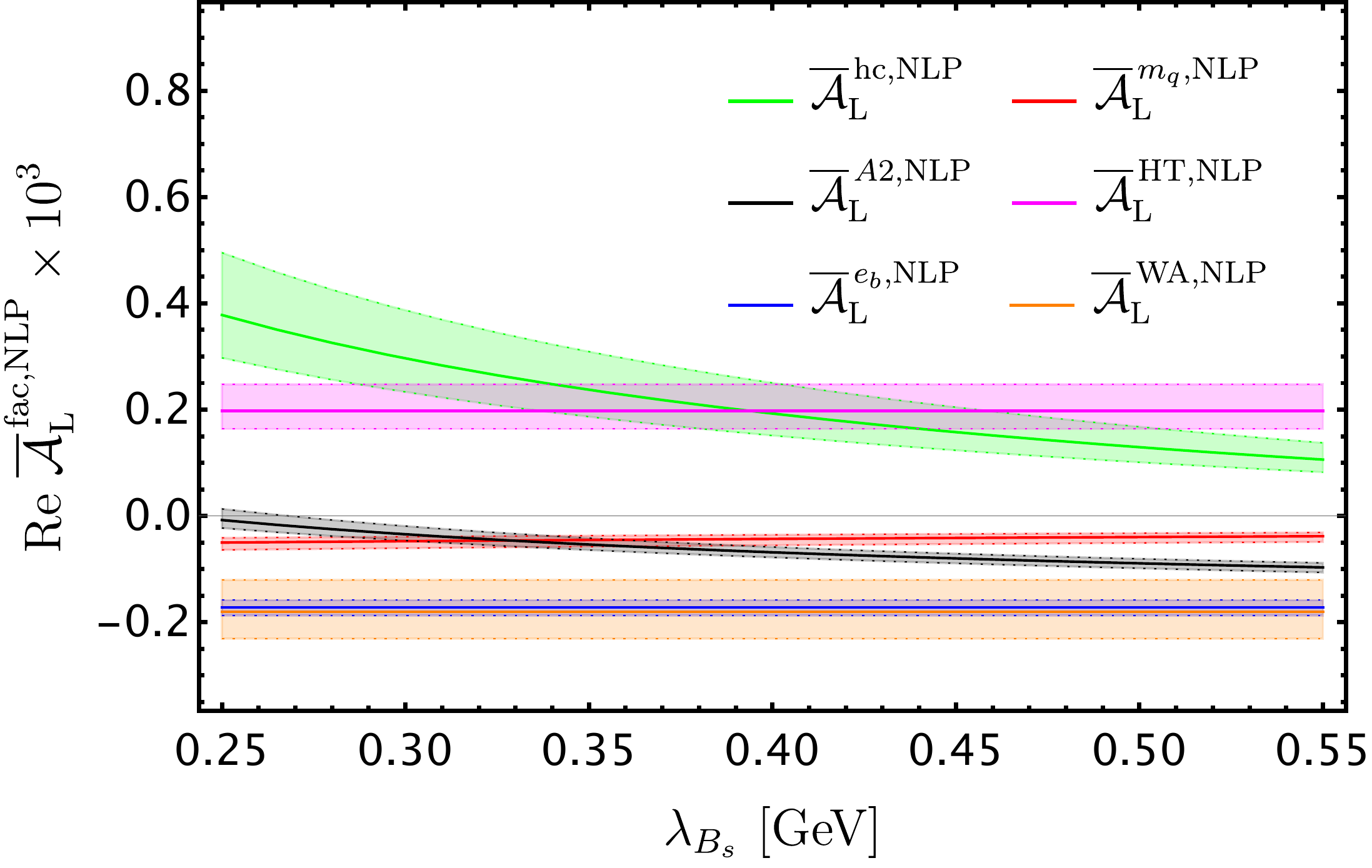}
\vspace*{0.1cm}
\caption{The leading power  and various NLP amplitudes in $\Bar{B}_s \to \gamma\gamma$ decays as a function of $\lambda_{B_s}$. Except for the strange quark mass contribution, the other amplitudes are from ref.\cite{Shen:2020hfq}.}
\label{fig: Breakdown of the full helicity amplitudes in Bsgg}
\end{center}
\end{figure}

Inputting all the  values of the parameters into the analytic formulae, we can obtain the numerical results for  various decay amplitudes and the phenomenological observables.
In what follows, we first present the numerical result of the strange quark mass effect in  the $\bar{B}_s \to \gamma\gamma $ decays. The  $\lambda_{B_s}$ dependence of the real part of the left-handed polarized amplitude from the leading power contribution, the NLP contribution without quark mass effect and the full result is shown in the left panel of  Fig. \ref{fig: Breakdown of the full helicity amplitudes in Bsgg}. It can be seen that the contribution from quark mass  is a  few percent of the full amplitude, which  is almost independent of the parameter $\lambda_{B_s}$. As a result, it plays a more important rule in the NLP contribution as $\lambda_{B_s}$ becomes large. The right panel of  Fig. \ref{fig: Breakdown of the full helicity amplitudes in Bsgg} exhibits the amplitudes from various sources of NLP contribution, and it is obvious that the cancellation between them highlights the contribution from quark mass term. We should also mention that the newly included term proportional to $m_s^2$ is numerically negligible, although it seems at the same power with the $m_s$ term.

\begin{figure}
\begin{center}
\includegraphics[width=0.6 \columnwidth]{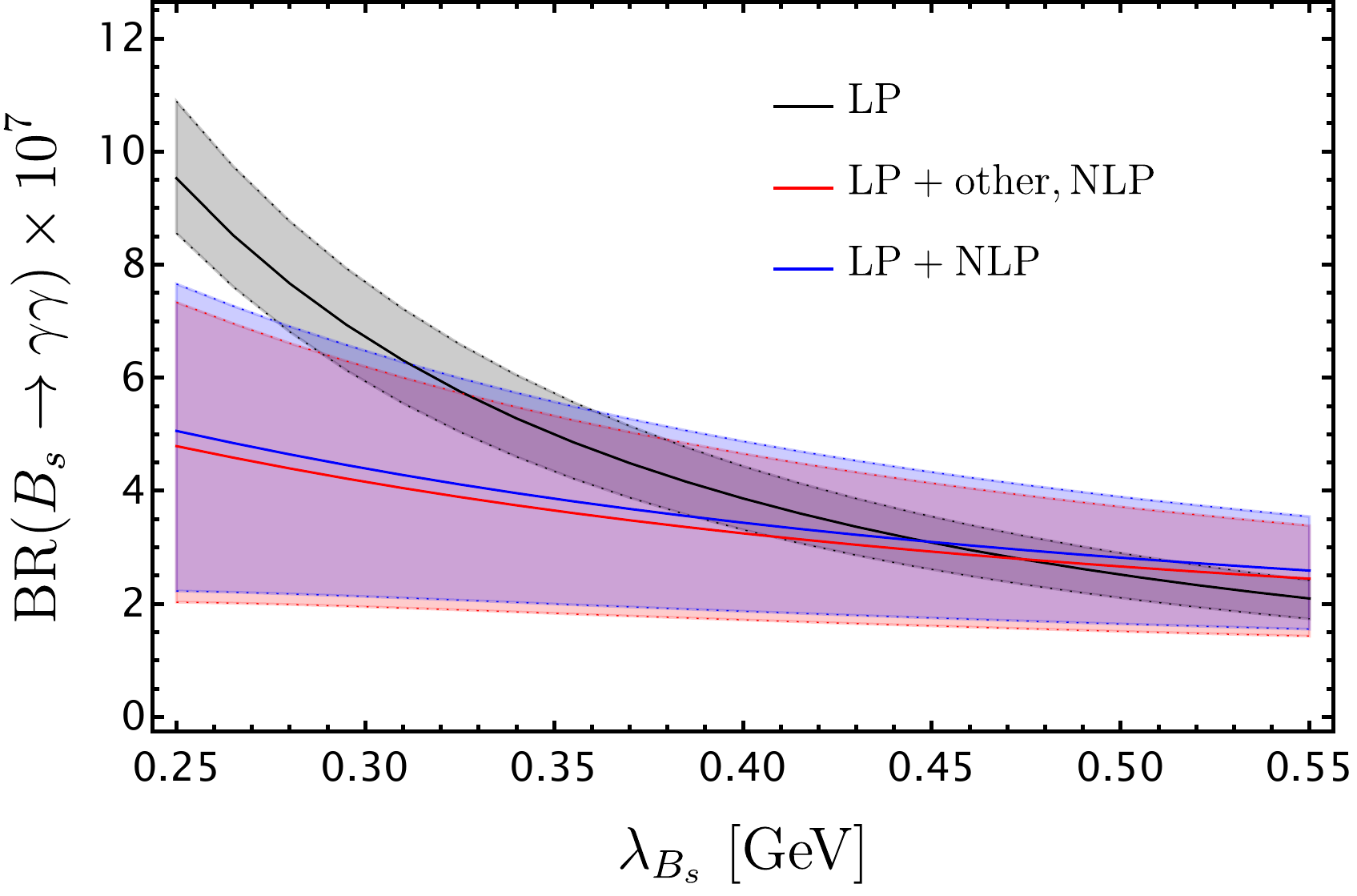}
\vspace*{0.1cm}
\caption{The branching ratio of $B_s \to \gamma\gamma$ decays  as a function of $\lambda_{B_s}$: (1) with only leading power contribution; (2) with NLP contributions except for strange quark mass term; (3) The complete contribution.}
\label{fig: breakdown of BR in Bsgg}
\end{center}
\end{figure}

The branching ratio of $B_s \to \gamma\gamma$ decay is given in Table~\ref{tab:  BR in Bsgg}, where the  uncertainty   is obtained by varying separate input parameters within their ranges and adding the   different uncertainties of the form factors in quadrature.   The main uncertainty is obviously from the parameter $\lambda_{B_s}$. From the central value we can see that the quark mass term can enhance the branching ratio about 6\%, which  deserves a reliable study if one intends to precisely determine the parameters in the standard model. In Fig. \ref{fig: breakdown of BR in Bsgg} we plotted the $\lambda_{B_s}$ dependence of the branching ratio of  $B_s \to \gamma\gamma$ decay, which can serve as a good method to determine $\lambda_{B_s}$. However, the accuracy of the determination of the parameter $\lambda_{B_s}$ is limited by the large theoretical uncertainty of
  the uncertainty of other  parameters. An optional method to improve the accuracy is to perform a global fit together with the other processes.

\begin{table}
\begin{center}
\caption{CP-averaged branching ratio of $B_s \to \gamma\gamma$ decays with uncertainty in unit of $10^{-7}$. }
\label{tab:  BR in Bsgg}
\renewcommand\arraystretch{1.8}
\begin{tabular}{c|c|c|c}
	\hline\hline
 Contributions &   { Central Value } &   { Total Error } &  {Error from }$\lambda_{B_{s}}$ \\
	\hline {leading power} & 3.87 & ${ }_{-1.85}^{+5.69}$ & ${ }_{-1.77}^{+5.66} $\\
	\hline {NLP} &3.25 & ${ }_{-1.73}^{+2.09}$ &$ { }_{-0.80}^{+1.54} $\\
	\hline   {NLP+quark mass} & 3.44 & ${ }_{-1.78}^{+2.17} $&$ { }_{-0.85}^{+1.62} $\\
	\hline\hline
\end{tabular}
\end{center}
\end{table}

\begin{figure}
\begin{center}
\includegraphics[width=0.4 \columnwidth]{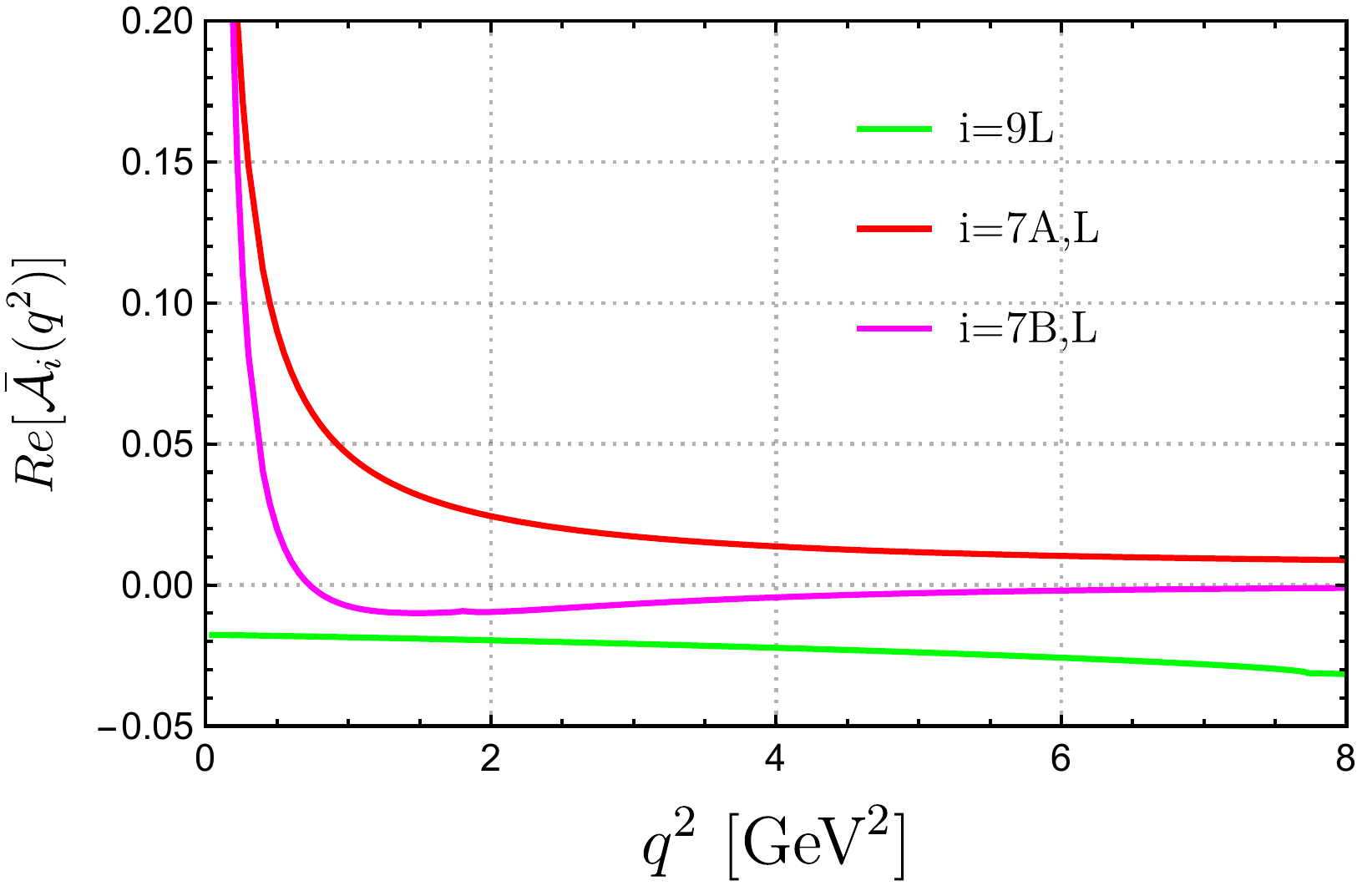}
\includegraphics[width=0.4 \columnwidth]{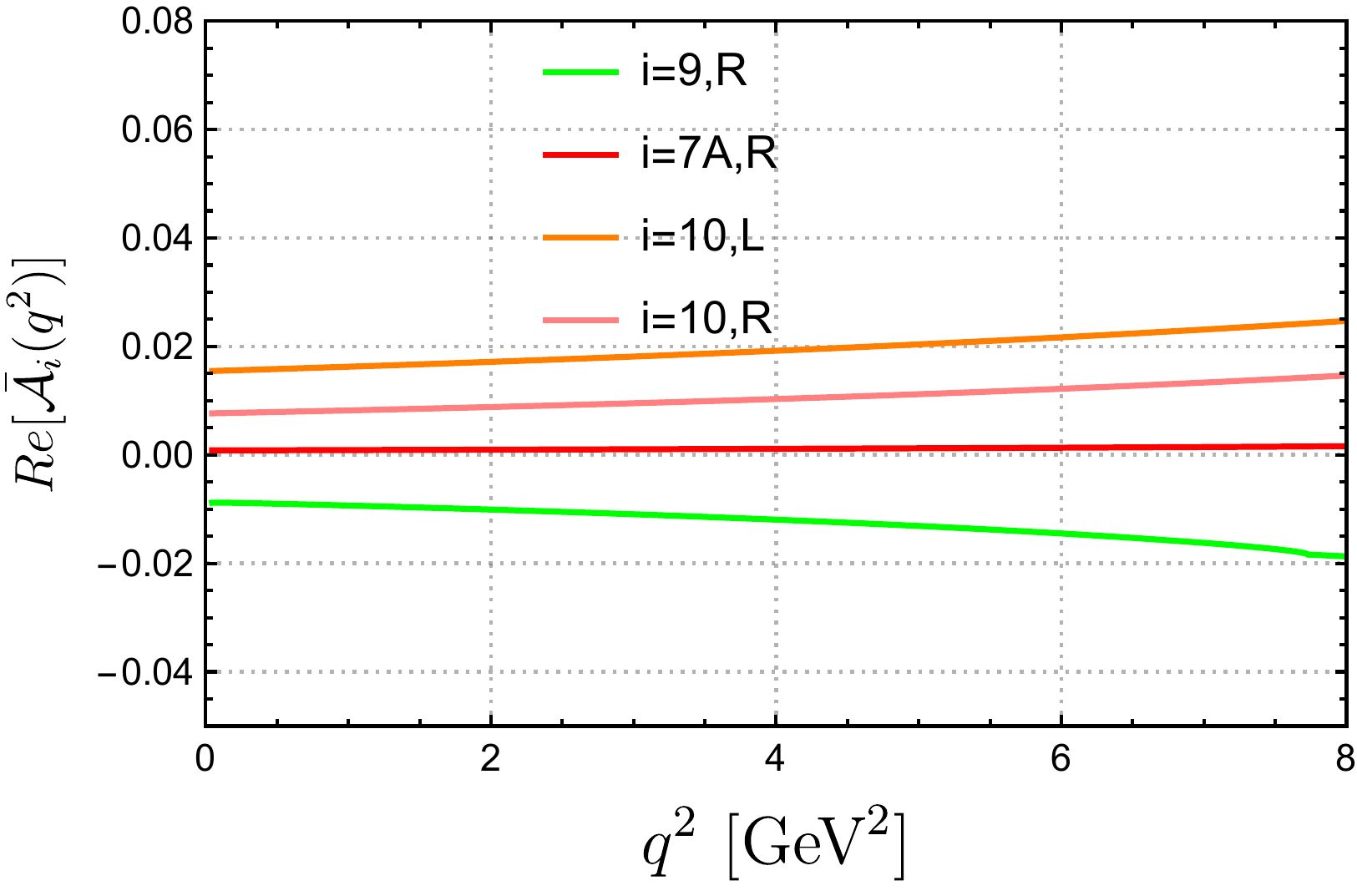}
\vspace*{0.1cm}
\caption{The  quark mass contribution to the real parts of $\bar{B}_s \to \gamma\mu^+{\mu}^- $ decay amplitude $\bar{A}_{\text{LV}}$ [left] and $\bar{A}_{\text{RV,RA,LA}}$ [right]  from operator $P_{9,10}$ and $P_{7}$   as a function of $q^2$. }
\label{fig: Breakdown of the full helicity amplitudes comparison}
\end{center}
\end{figure}

\begin{figure}
\begin{center}
\includegraphics[width=0.4 \columnwidth]{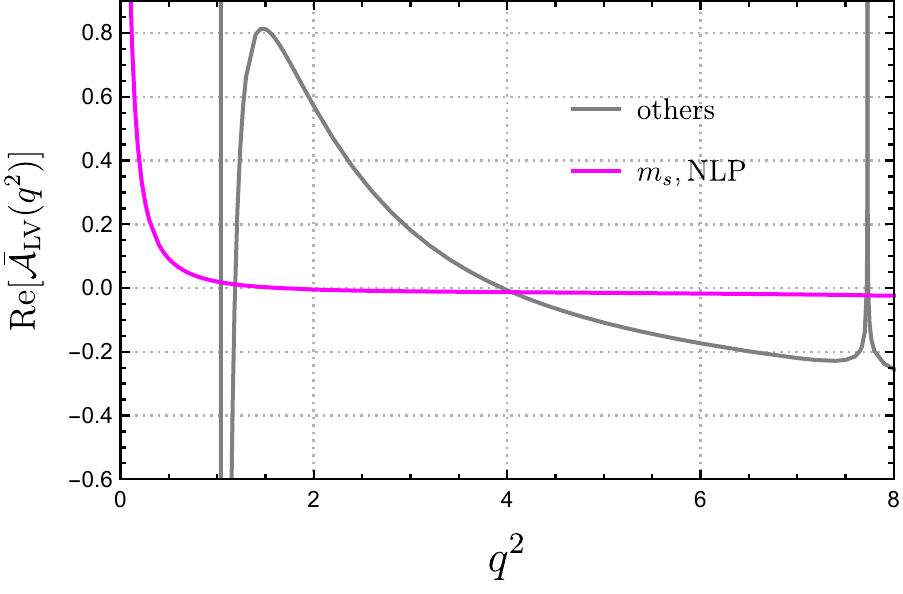}
\includegraphics[width=0.4 \columnwidth]{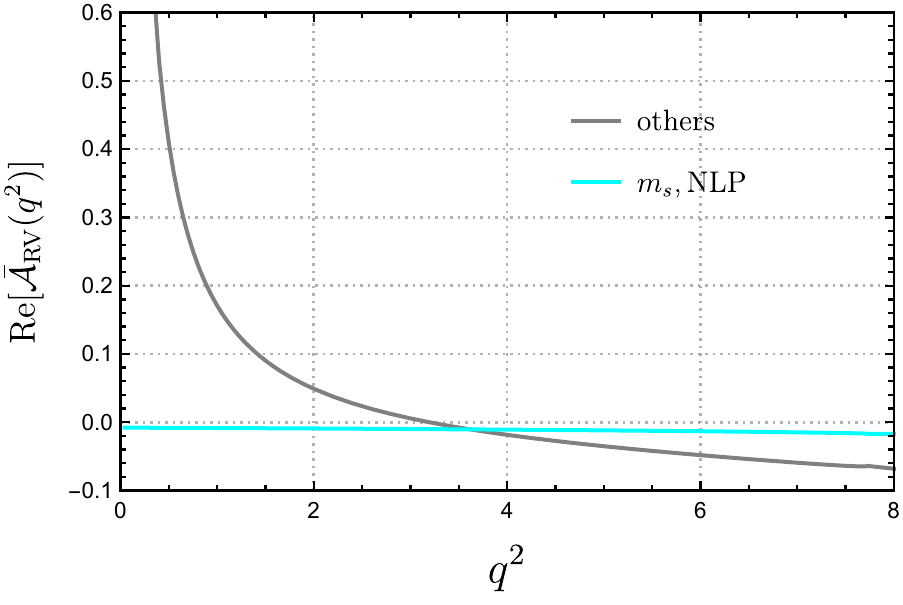}
\vspace*{0.1cm}
\caption{The    quark mass [violet, cyan] contribution to real parts of $\bar{B}_s \to \gamma\mu^+{\mu}^- $ decay amplitude $\bar{A}_{\text{LA}}$  (left) and $\bar{A}_{\text{RA}}$ (right)   from   operator $P_{7}$ and   $P_{9}$  as a function of $q^2$. For comparison the total results with $\phi(1020)$ resonance around $q^2\simeq 1$ GeV [gray] are also shown.}
\label{fig: Breakdown of the full helicity amplitudes O7O9}
\end{center}
\end{figure}

The strange quark mass effect in  the $\bar{B}_s \to \gamma\mu^+{\mu}^- $ decay is a bit more complicated since several operators can contribute besides the electromagnetic penguin operator, and the right-handed polarized amplitude can also receive   corrections from the strange quark mass effect.  The size of quark mass term induced by the operators $P_{7,9,10}$ are collected in Fig.~\ref{fig: Breakdown of the full helicity amplitudes comparison} as a function of $q^2$. We can see that almost all the effects of axial-vector currents or right-helicity amplitudes come from semileptonic operators $P_{9,10}$. It is clear that    there exists cancellation between different operators, which reduces the total quark mass term  contribution.
In the amplitude level, the ratio between the quark mass term and the full amplitude is close to that of the $\bar{B}_s \to \gamma\gamma$ decays. We exhibit relative size between the quark mass effect and the full amplitude in   Fig.~\ref{fig: Breakdown of the full helicity amplitudes O7O9}   induced by   $P_{7,9}$ operators. In this figure, the total results with $\phi(1020)$ resonance contribution around $q^2\simeq 1$ GeV [gray] are also shown. It is easy to see that the mass term contribution in  the $\bar{B}_s \to \gamma\mu^+{\mu}^- $ decay is negligibly small except a larger contribution at very small $q^2$. This sizable contribution is from operator $P_7$, which can be seen from Fig. \ref{fig: Breakdown of the full helicity amplitudes comparison}.

\begin{figure}
\begin{center}
\includegraphics[width=0.5\columnwidth]{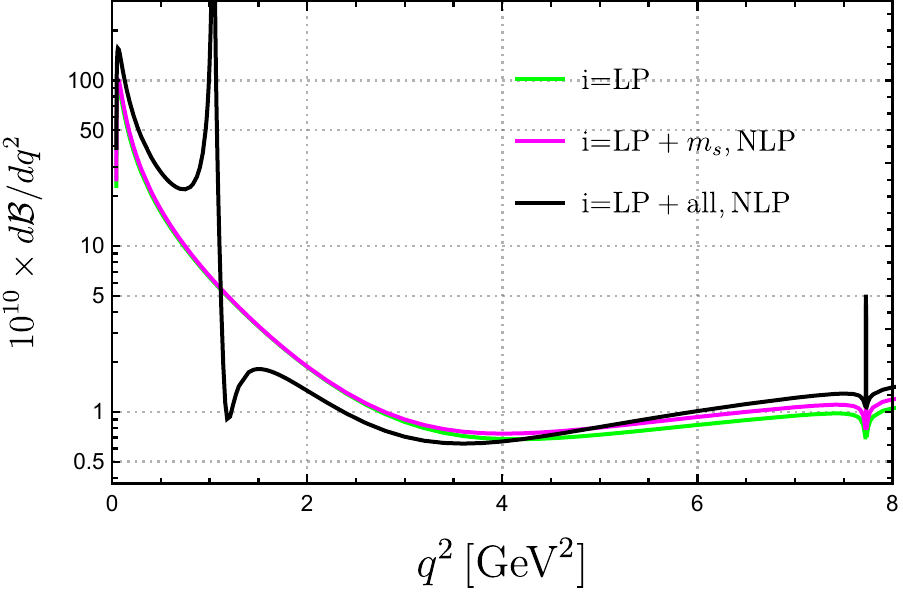}
\vspace*{0.1cm}
\caption{The CP-averaged differential branching ratio $d \mathcal{B} / d q^{2}$ distributions for $B_{s} \rightarrow \gamma \mu^+  {\mu}^-$ decay with and without the mass term contribution.}
\label{fig: Breakdown of BR gall}
\end{center}
\end{figure}

\begin{table}
\begin{center}
\caption{Integrated branching ratios (in unit of $10^{-9}$) of $
B_s \to \gamma \mu^+{\mu}^-$ decay with and without quark mass effect. }
\label{brgll}
\renewcommand\arraystretch{1.6}
\begin{tabular}{c|c|c}
	\hline\hline
	    { Region of} ~$q^2$ & $[4m_{\mu}^2,6.0]${GeV}$^2$&  {$[4m_{\mu}^2,8.0]$ {GeV}$^2$}\\
\hline
\hline   {without}~$m_s$ &$12.43  _{-1.93}^{+3.83}$&$12.74 _{-2.08}^{+4.15}$  \\
\hline   {with} ~$m_s$ & $12.73  _{-1.93}^{+3.83} $&$ 13.06_{-2.08}^{+4.15} $\\
\hline\hline
\end{tabular}
\end{center}
\end{table}

The differential branching ratio of $B_{s} \rightarrow \gamma \mu^+  {\mu}^-$ decay with respect to $q^2$ is plotted in Fig. \ref{fig: Breakdown of BR gall}, where we have included the on-shell hadronic state contribution in order to compare with the future data. The quark mass effect is negligible at small $q^2$ region, due to the large hadronic resonance contribution.
The integrated branching ratios are listed in Table \ref{brgll}, where we have considered two integration regions $ [4m_{\mu}^2,6.0]${GeV}$^2$ and $[4m_{\mu}^2,8.0]$ {GeV}$^2$ of   invariant mass of the lepton pair. The  uncertainty from s-quark mass term is so insignificant that it is not taken into account in the total error from \cite{Beneke:2020fot}.
The results in this table indicate that the quark mass effect is less important in $B_s \to \gamma\mu^+ {\mu}^- $ decay than that in the $B_s \to \gamma\gamma$ decay. This is mainly due to the inclusion of the hadronic state contribution at small $q^2$, which significantly enhance the total branching ratio.


\section{Summary}

The power suppressed contributions play an important role in the    radiative decays $B_{d, \, s} \to \gamma \gamma$ and radiative leptonic decays $B_{d, \, s} \to \gamma \ell\bar{\ell}$. Some of them are factorizable and can be investigated using factorization approach, however, most of them   can not be factorized due to the emergence of the endpoint singularity. Therefore, one must find some special methods to deal with them.  The contribution from quark mass term is nonfactorizable  in the $B_{ \, s} \to \gamma \gamma$ as well as $B_{ \, s} \to \gamma \ell\bar{\ell}$ decays. In the previous study it is parameterized  in a model dependent way. In order to reduce the model dependence and improve the theoretical precision, we revisit  this NLP contribution with a QCD-inspired approach, namely the dispersion approach. In this approach, we introduce the $B_s \to V$ form factors instead of the arbitrary momentum cut off to deal with the endpoint singularity, therefore, it is more predictive. We have presented the analytic expression of the quark mass contribution  in the $B_{ \, s} \to \gamma \gamma$ and $B_{ \, s} \to \gamma \ell\bar{\ell}$ decays in the new approach, together with a new term that is missed in the previous study.
 
The numerical results of the NLP contribution to  the $B_{ \, s} \to \gamma \gamma$ and $B_{ \, s} \to \gamma \ell\bar{\ell}$ decays from the strange quark mass effect have also been given. In the $B_{ \, s} \to \gamma \gamma$ decay, the strange quark mass term can give rise to about $6\%$ contribution relative to the total amplitude, which makes sense if this process is employed to determine the parameters in the standard model. The terms proportional to $m_s^2$ which has been omitted in the previous study is numerically very small.  The strange quark mass contribution to the $B_ { \, s} \to \gamma \ell\bar{\ell}$ decays is relatively small, due to the cancellation between the contributions from different operators and the enhancement of large hadronic resonance contribtion.  The uncertainty of the input parameters is sizable, which renders us from a more accurate prediction so far.  Anyway, with our improved theoretical method,  we can arrive at more precise predictions, if future experiments can      precisely constrain the hadronic parameters.

\section*{Acknowledgement}

We thank Yan-Bing Wei and Hao-Yi Ci for some useful contributions. 
This work was supported in part by the National Natural Science Foundation of China under
Grant No.~12175218 and 12070131001 and  the National Key Research and Development Program of China under Contract No.2020YFA0406400.
Y.L.S also acknowledges the Natural Science Foundation of Shandong province with Grant No. ZR2020MA093.

\end{document}